\documentclass[aps,prb,twocolumn,floatfix,floats]{revtex4}
\usepackage{graphicx,latexsym,t1enc}

\begin{document}

\title{Non-adiabatic current generation in a finite width semiconductor ring}

\author{Vidar Gudmundsson}
\altaffiliation[Permanent address: ]{Science Institute, University of Iceland, 
        Dunhaga 3, IS-107 Reykjavik, Iceland}
\affiliation{Physics Division, National Center for Theoretical Sciences,
        P.O.\ Box 2-131, Hsinchu 30013, Taiwan}
\author{Chi-Shung Tang}
\affiliation{Physics Division, National Center for Theoretical Sciences,
        P.O.\ Box 2-131, Hsinchu 30013, Taiwan}
\author{Andrei Manolescu}
\affiliation{Science Institute, University of Iceland, 
             Dunhaga 3, IS-107 Reykjavik, Iceland}

%
%

\begin{abstract} 
We consider a model of a semiconductor quantum ring of finite 
width in a constant
perpendicular magnetic field. We show how a current of the same order
as the persistent current can be generated non-adiabatically 
by a short intensive pulse in the Tera-Hertz regime.
\end{abstract}


\maketitle
%
Time-dependent or radiation induced phenomena in various 
mesoscopic electron systems have caught the attention of 
experimental\cite{Mani02:646} 
and theoretical\cite{Tang01:353} groups recently,
just to mention two cases of a long list. 
The generation of a current by a slowly oscillating harmonic magnetic flux  
in a strictly one-dimensional (1D) ballistic quantum ring
with a spinless Luttinger liquid has been investigated by 
Moskalets\cite{Moskalets00:349} finding that the AC part depends both
on the driving frequency and the temperature, but the DC part of the
current is only very sensitive to the driving amplitude.   
More recently Moskalets and B{\"u}ttiker have discovered a strong
parity effect in a 1D quantum wire excited by a harmonic delta-function 
potential.\cite{Moskalets02:0207258} 
They find the current in a ring with an odd number
of spinless electrons to be diamagnetic, exactly like in a pure ring 
without an excitation, but 
in a ring with an even number the current oscillates in sign with a large
amplitude and with a small period compared to the unit flux quantum
$\Phi_0=h/(ec)$.  

Inspired by the orbital current modes found in elliptical quantum 
dots by Serra et al.,\cite{Serra99:13966} and the large effects of a nonlinear
excitation of quantum dots,\cite{Puente01:235324} 
we present here an alternative method to generate DC currents in quantum rings 
nonadiabatically with a strong and short THz pulse. In our case the finite
width of the quantum ring is essential and the current generation does
rely on a change in the many-electron structure of the system caused
by the THz pulse. 

The equilibrium persistent currents and the magnetization of
quantum rings of finite width has been studied by Tan and 
Inkson,\cite{Tan99:5626} and Vasile et al.,\cite{Vasile02:xx} 
and their low-intensity 
far-infrared absorption has been measured by Lorke 
et al.,\cite{Lorke00:2223} and calculated by several 
groups.\cite{Zaremba96:R10512,Ingibjorg99:16591,Puente01:125334}  

To calculate the time evolution of the system requires the knowledge
of the initial state, its ground state, at a particular time $t=t_0$,
which within a suitable mean field approximation is described by 
\begin{equation}
      H|\alpha ) = \left(  H_0 + H_\sigma + V_{c} + H_{\mathrm{int}} 
      \right) |\alpha ) = \varepsilon_\alpha |\alpha ) ,
      \label{Kohn_Sham} 
\end{equation}
where $H_0$ is the Hamiltonian for the 2DEG, that is subject to 
the perpendicular magnetic field ${\bf B}=B{\bf\hat z}$, and confined by the
parabolic potential $V_{\mathrm{par}}(r)=m^*\omega_0^2r^2/2$.
We modify the radial confinement with a central potential hill, 
$V_{c}(r) = V_0\exp{(-\gamma r^2)}$, creating the confining potential 
$V_{\mathrm{conf}} = V_{\mathrm{par}} + V_{c}$ in order to make the system
ring like. The spin-dependent part of the Hamiltonian is $H_\sigma =
(1/2)g^*\mu_BB\sigma_z$, where $g^*$ is an effective $g$-factor, $\mu_B$ the
Bohr magneton, and $\sigma_z=\pm 1$ for the two spin projections.

The Hamiltonian $H_0$ defines a natural length scale, $a = l\sqrt{\omega_c/\Omega}$,
in terms of the magnetic length, $l = \sqrt{\hbar c/(eB)}$, the cyclotron
frequency $\omega_c = eB/(m^*c)$, and the characteristic frequency,
$\Omega = \sqrt{\omega_c^2 + 4\omega_0^2}$. 
The eigenstates $|\alpha\rangle$, with $\alpha = (M,n_r)$, 
of the Fock Darwin\cite{Fock28:446}
Hamiltonian $H_0$ are used as a mathematical basis in which the eigenstates
of $H$, $|\alpha )$, are expanded in, and in which (\ref{Kohn_Sham}) is transformed
into a matrix eigenvalue equation. No restriction is put on the angular
symmetry of the system. A suitable truncation of the
basis $\{|\alpha\rangle\}$, that can be varied to improve the accuracy,
leads to a numerically tractable eigenvalue problem.

At $t=t_0$ the quantum ring is radiated by a short THz pulse  
making the Hamiltonian of the system time-dependent,
$H(t)=H+W(t)$, with
\begin{eqnarray} 
      W(t) &=& V_t r^{|N_p|}\cos{(N_p\phi)}\exp{(-sr^2-\Gamma t)}\nonumber\\ 
           &{\ }&\sin{(\omega_1t)}\sin{(\omega t)}\theta (\pi -\omega_1t) ,
\label{Wt}
\end{eqnarray}
where $\theta$ is the Heaviside step function.
The parameters, $s$, $\omega_1$, $\omega$, and $N_p$ shall be
explained below. Since the radiation is of arbitrary strength 
we resume to the task of solving the equation of motion for 
the density operator
\begin{equation}
      i\hbar \frac{d}{dt}{\rho}(t) = [H + W(t),\rho (t)] .
\label{eq_motion}
\end{equation}
The structure of this equation is inconvenient for numerical evaluation
so we resort instead to the time-evolution operator $T$, defined by
$\rho (t) = T(t)\rho_0T^+(t)$, which has the simpler equation of motion
\begin{eqnarray}
      i\hbar\dot T(t)   &=& H(t)T(t)\nonumber\\    
     -i\hbar\dot T^+(t) &=& T^+(t)H(t)  .
\label{Teq}     
\end{eqnarray}
We discretize time and use the Crank-Nicholson algorithm for the 
time-integration with the initial condition, $T(0)=1$. 
This is performed in the truncated
Fock-Darwin basis $\{|\alpha\rangle\}$ for the corresponding
matrix version of the Eqs.\ (\ref{Teq}). 

As a matter of convenience, we use the time-dependent orbital magnetization
${\cal M}_o(t)$ to quantify the AC and DC currents induced by $W(t)$. 
The magnetization operator $\mathbf{M}_o = (1/2c)\mathbf{r}\times\mathbf{j}$ 
together with the definition for the current, 
$\mathbf{j} = -e\dot\mathbf{r} = (-ie/\hbar )[H(t),\mathbf{r}]$, give 
in terms of the density matrix in the Fock-Darwin basis 
$\{ |\alpha\rangle\}$ the magnetization  
\begin{eqnarray}
      \lefteqn{{\cal M}_o(t) = -\frac{e}{2c}\mathrm{tr}\{ (\mathbf{r}
                               \times\dot\mathbf{r})\cdot\hat\mathbf{z}\;\rho (t)\}
                             =   - \frac{ie}{2\hbar c}
                      \sum_{\delta\tau\beta\gamma}\rho_{\tau\delta}(t)\nonumber}\\
      & &\{ \langle\delta |x|\beta\rangle 
      (  \langle\beta |H(t)|\gamma\rangle\langle\gamma |y   |\tau\rangle
        -\langle\beta |y   |\gamma\rangle\langle\gamma |H(t)|\tau\rangle)\nonumber\\  
      & & -\langle\delta |y|\beta\rangle
      ( \langle\beta |H(t)|\gamma\rangle\langle\gamma |x   |\tau\rangle
      - \langle\beta |x   |\gamma\rangle\langle\gamma |H(t)|\tau\rangle )\}
\end{eqnarray}

In this communication, for reasons of clarity and space, 
we only report on results for non-interacting electrons.
The inclusion of the Coulomb interaction in a LSDA (Local Spin-Density
Approximation) manner only changes the results here quantitatively, 
and shall be reported elsewhere. It should though be stated here, 
since it is also relevant to the noninteracting model, that 
the accuracy and the fine-tuning of the LSDA model have been tested for the
ground state with comparison to the exact results calculated for a 
circular parabolically confined quantum dot,\cite{Pfannkuche93:2244} 
and in the time-evolution by convincing us that the model can reproduce
the Kohn results for the absorption of a Far-infrared 
radiation.\cite{Maksym90:108,Gudmundsson91:12098}

In order to model a ring-shaped 2DEG with a number of electrons, 
$N = 1 - 13$, in a GaAs sample with vanishing 
electron density in its center we select the parameters, 
$\hbar\omega_0 = 3.37$ meV, $V_0 = 30$ meV, and $a^2\gamma =1.0$
in the center part of the confinement potential $V_c$.
We assume $g = 0.44$, and $m^* = 0.067m$. As this small
effective mass enhances the orbital magnetization of the
2DEG with a factor of $14.9$ with respect to the
spin magnetization we shall ignore the latter here. 
Moreover, we are here merely using the orbital magnetization
of the system as a measure of the persistent currents in it.
For the radiation pulse (\ref{Wt}) of just over 3 ps, we select
$\Gamma = 2$ ps$^{-1}$. The envelope frequency corresponds to
$\hbar\omega_1 = 0.658$ meV, and the base frequency is 
$\hbar\omega = 2.63$ meV. 
A conventional dipole radiation is described by $|N_p|=1$ in
Eq.\ (\ref{Wt}), but we shall also consider here higher multipoles
with $|N_p|=2,3$, or the monopole, $N_p=0$, which could only be produced by
specially patterned microwave antennae or gate fingers.
For $N_p = 0$ we use $sa^2=1.0$, but otherwise $s = 0$.
We use a basis set with $M^{\mathrm{max}} = 12$, and
$n_r^{\mathrm{max}}\in [4,8]$.  

\begin{figure}[htbp!]
\begin{center}
\includegraphics[width=8.6cm]{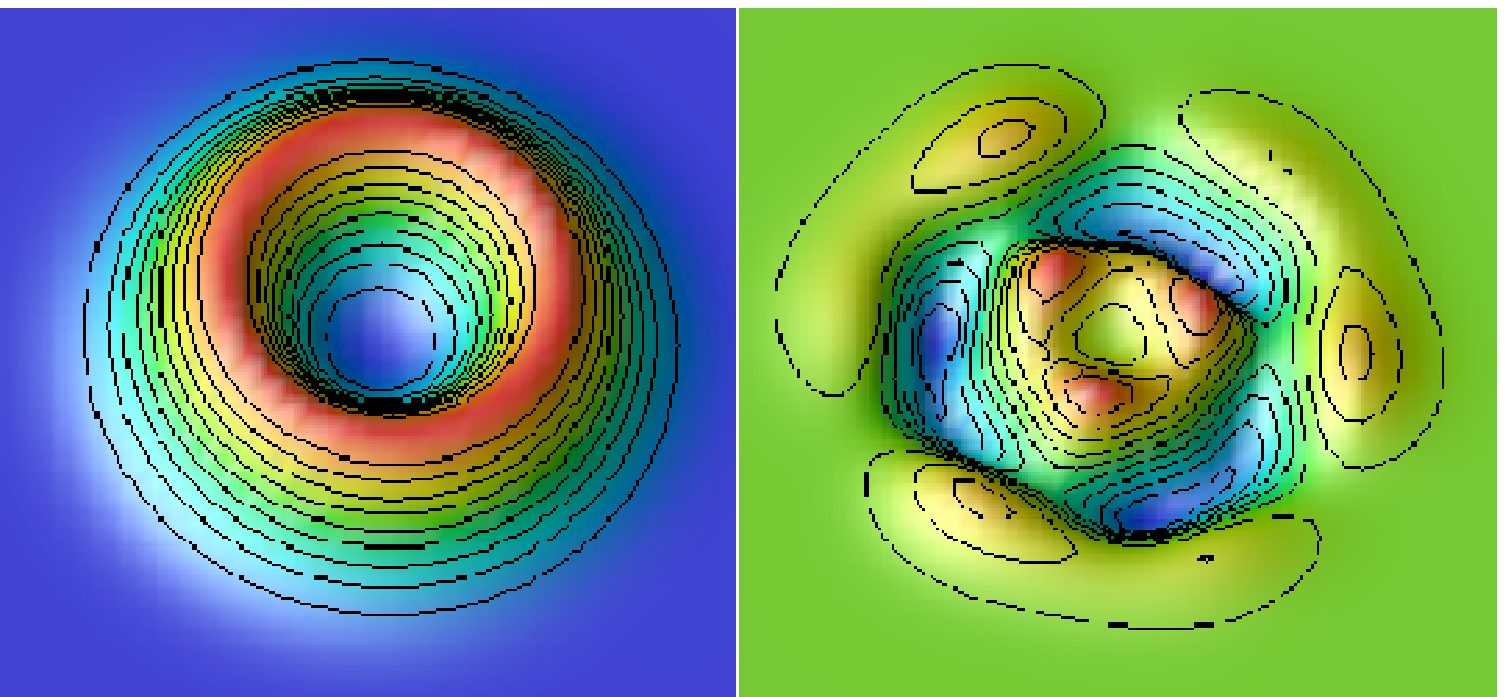}
\end{center}
\caption{The electron density (left), and the induced density (right) at some
         later point in time. $B = 0.6$ T, $|N_p| = 3$, 
         $V_ta^{|N_p|} = 1.0$ meV, $N = 12$, and $T = 1$ K.}
\label{Dens}
\end{figure}

\begin{figure}[htbp!]
\begin{center}
\includegraphics[width=8.6cm]{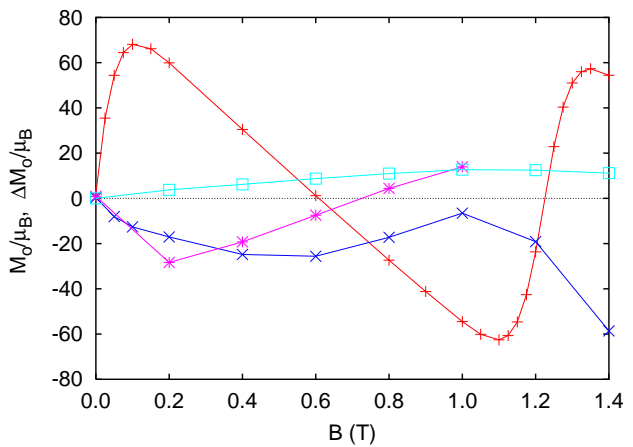}
\end{center}
\caption{The equilibrium magnetization ${\cal M}_o(0)$, $(+)$, and the 
        averaged time-dependent change in the magnetization $\Delta{\cal M}_o
        = \langle{\cal M}_o(t) - {\cal M}_o(0) \rangle$ 
        after the radiation pulse has vanished for $|N_p| = 1$ $(\Box)$,   
        $|N_p| = 2$ $(*)$, and $|N_p| = 3$ $(\times)$. For $|N_p| = 2$
        and 3 $V_ta^{|N_p|} = 1.0$ meV, and for $|N_p| = 1$ 
        $V_ta^{|N_p|} = 2.0$ meV. For $|N_p| = 1$, and $|N_p| = 2$ 
        the curves have been magnified by a factor of 10.
        $N = 12$, $T = 1$ K, 
        and other parameters are in text.}
\label{M_N12}
\end{figure}

We integrate the equations for the time-evolution operator
(\ref{Teq}) for a time interval of 40 ps using an increment
in the range of 0.002 - 0.01 ps and evaluate ${\cal M}_o(t)$
at each step. The Fourier transform of the time series in the 
interval 4 - 40 ps, when the external excitation has been turned off, 
yields information about the average 
magnetization in this interval, $\langle{\cal M}_o\rangle$, and the oscillations
of the time-dependent part. The external pulse (\ref{Wt}) always 
produces a radial force on the 2DEG starting radial oscillations
of its density in the confining potential $V_{\mathrm{conf}}$. 
For all the values of $N_p$ considered 
here this oscillation is characterized by one or few narrow resonances
close to 7 meV, even though the driving frequency was 
$\hbar\omega = 2.63$ meV.  In an interacting system this frequency
is modified by the Coulomb interaction. In Fig.\ \ref{Dens}    
we show the equilibrium density for 12 electrons in a ring together
with the induced density at a later time for $N_p = 3$.
%
%
%
On top of the radial oscillation there is a clear angular oscillation
visible in Fig.\ \ref{Dens}. The average magnetization $\langle{\cal M}_o\rangle$
after the excitation pulse has been turned off is {\em different} 
from the magnetization
of the corresponding ground state ${\cal M}_o(0)$ as shown in 
Fig.\ \ref{M_N12}.
%
%
%
%
Moreover, pulse strength here is large enough to increase the average size 
of the system, just as has been observed in highly excited quantum 
dots.\cite{Puente01:235324} We are in the nonlinear regime of excitation and
have nonadiabatically pumped energy into the system. As a result the 2DEG
is now in an excited state with a different persistent current.
This conclusion is supported by comparing the density matrix before and
after the excitation. 

We can neither generate a nonzero current this way at vanishing magnetic field,
$B=0$, nor with a pure radial excitation, $N_p=0$ at a finite magnetic field.
The reason for the $|N_p| = 3$ excitation to be more effective here than the 
$|N_p| = 2$ excitation is partly due to their different radial strength, and partly
due to the stronger angular modulation for the $|N_p| = 3$ case. 
The $|N_p=1|$ excitation is different in the sense that it also causes a large
oscillation of the center of mass of the system, a dipole oscillation 
absent in the other cases. The change $V_t\rightarrow -V_t$ is ineffective
since it can always be canceled, for symmetry reasons, by a 
fixed-angle rotation of the
2DEG, the numerical model also reproduces this result. 

\begin{figure}[htbp!]
\begin{center}
\includegraphics[width=8.6cm]{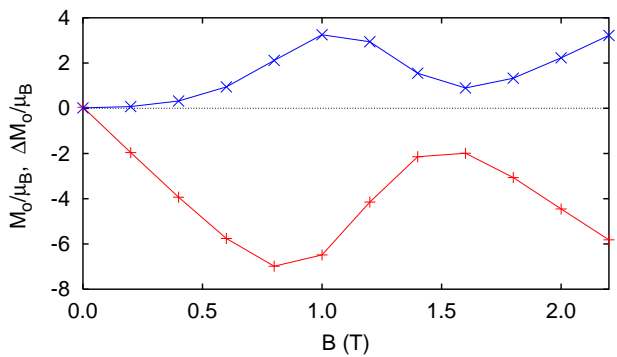}
\end{center}
\caption{The equilibrium magnetization ${\cal M}_o(0)$, $(+)$, and the 
        averaged time-dependent change in the magnetization $\Delta{\cal M}_o$ 
        after the radiation pulse has vanished for $|N_p| = 3$ $(\times)$.
        $V_ta^{|N_p|} = 1.0$ meV, $N = 1$, and $T = 1$ K.}
\label{M_N01}
\end{figure}

\begin{figure}[htbp!]
\begin{center}
\includegraphics[width=8.6cm]{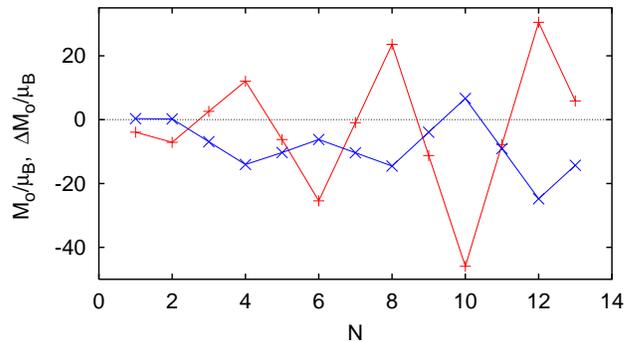}
\end{center}
\caption{The equilibrium magnetization ${\cal M}_o(0)$, $(+)$,
         and the averaged time-dependent change in the magnetization 
         $\Delta{\cal M}_o$
         after the radiation pulse has vanished for $|N_p| = 3$ $(\times)$ as
         functions of the number of electrons $N$. $V_ta^{|N_p|} = 1.0$ meV, 
         $B = 0.4$ T, and $T = 1$ K.}
\label{M_N}
\end{figure}

\begin{figure}[htbp!]
\begin{center}
\includegraphics[width=8.6cm]{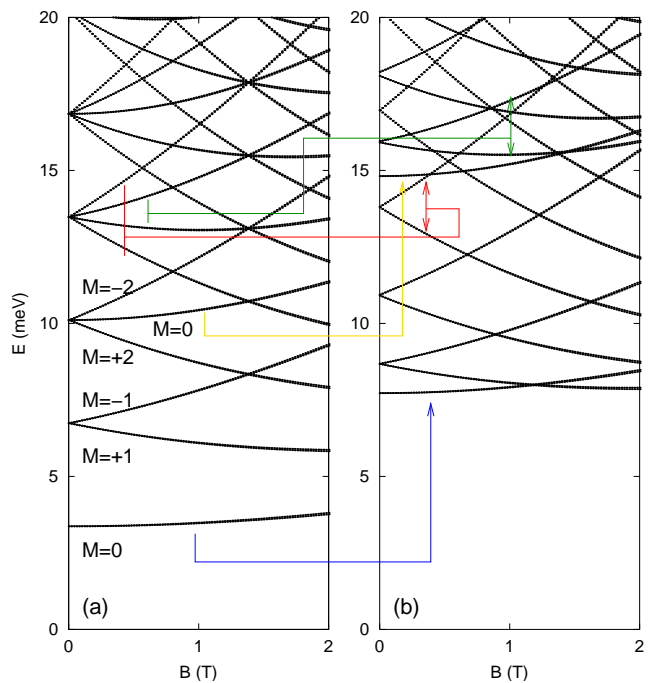}
\end{center}
\caption{The noninteracting single-electron energy spectrum
         for a quantum dot, $V_0 = 0$ meV (a), and a quantum
         ring with $V_0 = 30$ meV (b). Arrows connect some low 
         energy states with the same values of angular quantum number
         $M$. $T = 1$ K, see other parameters in text.}
\label{E_DF}
\end{figure}

\begin{figure}[htbp!]
\begin{center}
\includegraphics[width=8.6cm]{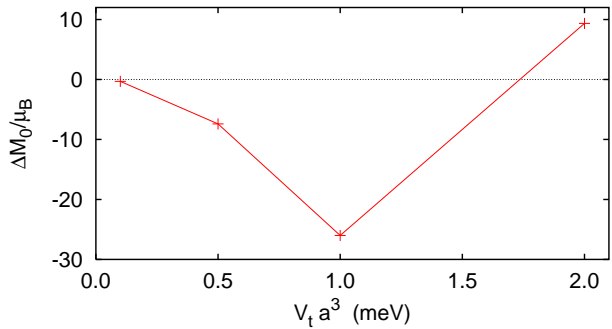}
\end{center}
\caption{The time-dependent change in the magnetization $\Delta{\cal M}$
         after the radiation pulse has vanished for $|N_p| = 3$ $(+)$ as
         a function of the pulse strength $V_t$. $T = 1$ K.}
\label{M_Vt}
\end{figure}

The current generated depends strongly on the number of electrons in the ring
as can be seen by comparing the results for one electron in Fig.\ \ref{M_N01}
to the ring with 12 electrons, Fig.\ \ref{M_N12}. 
%
%
%
%
A certain parity effect can be observed in the sign of ${\cal M}_o(0)$,
or the direction of the induced persistent current, see Fig.\ \ref{M_N}.
%
%
%
%
%
The magnetization in the ground state ${\cal M}_o(0)$ assumes a maximum value 
for an even number of noninteracting electrons with spin and is much 
smaller for an odd number.
This behavior can be understood in terms of the quasi periodicity of the
energy spectrum for a low magnetic field, see Fig.\ \ref{E_DF}(b).
%
%
%
%
Each energy level is almost spin degenerate and, for example, the change from 6
to 8 electrons changes the system from diamagnetic to paramagnetic as can
be guessed from the slope of the highest occupied level. In case of 7 electrons
the 2DEG pertains both qualities resulting in a small net magnetization.      
The induced magnetization $\langle{\cal M}_o\rangle$, on the other hand is in antiphase
with these oscillations. This can only be understood in terms of a changed 
character of the excited 2DEG. If the excitation is not too large then the
system gains through it more of the character of the lowest unoccupied state,
that fulfills the selection rule $|\Delta M| = |N_p|$.
Thus, a still stronger excitation might be able to reverse this trend, as
can be confirmed in Fig.\ \ref{M_Vt}, where an increase in the excitation
strength changes the 2DEG from diamagnetic to paramagnetic.  
%
%
%

By inspecting the diagonal elements of the ground state density matrix for the 
case of $N = 12$, and $B = 0.4$ T it can be seen that there is a strong overlap
with the 5 basis states with $n_r = 0$, and $M = -2, -1, +1, +2, +3$ and a partial
overlap with $(M,n_r) = (0,0)$ and $(0,1)$ reflecting the center potential hill
in the system, that pushes the low lying $M=0$ states to higher energy
as can be seen in Fig.\ \ref{E_DF}. With the center potential hill present
the states can not be assigned the radial quantum number $n_r$ anymore.
After the system has been excited with a $V_ta^3 = 1$ meV pulse
the overlapping with the states $(3,0)$, $(0,0)$, $(0,2)$, $(0,1)$, 
and $(1,1)$ continuously changes the most, reflecting a strong radial oscillation
with a smaller angular mode. The excited state does thus include $M$-states
in a different ratio causing a changed persistent current pattern. 
The occupation of still higher $M$-states needs still stronger excitation.  

Although the change in the average persistent current is a general
feature of the strong excitation described here,
clearly the antiphase found in Fig.\ \ref{M_N} is not a general feature 
of the system, but rather the relation between the ${\cal M}_o$ and
$\langle{\cal M}_o\rangle$ depends strongly on the relevant selection rule, 
$|\Delta M| = |N_p|$, and the
energy spectrum, which depends strongly on the
magnetic flux and is not simply periodic in the flux
for a ring of finite width, see Fig.\ \ref{E_DF}(b). 

We have demonstrated here that the persistent current, or its 
manifestation in the time-dependent magnetization, of a 
quantum ring of a finite width can be controlled by a strong external 
excitation pulse by inflicting a nonadiabatic change upon the many-electron 
state of the system. This current generation or change in the persistent
current can thus not be described by a linear response theory.
Due to the fast response of the system, at the level of excitation  
applied here, the change in its state occurs almost entirely during the time
the excitation pulse is applied to the 2DEG. The average magnetization
$\langle{\cal M}_o\rangle$, describing the new persistent current in the
system is evaluated after these switch-on effects have died down.    
Eventually the energy pumped into the system by the excitation pulse will
dissipate. We make no attempts to describe the dissipation process
and various possible channels thereof in the present model, but
anticipate the lifetime of the excited state - the collective mode -
to be long enough to allow for several cycles of oscillations before
at least the energy is carried away by electromagnetic radiation.

We have tested the results communicated here within a Density-Functional Theory
approach using the LSDA and found similar qualitative results for the limited 
values of the system parameters tested. It should be emphasized that the
neglect of the Coulomb interaction here is done consistently in the ground and
the excited state by integrating directly the equation of motion, something
that is not easy to perform within a linear response theory without risiking 
a violation of some conservation laws.       

\begin{acknowledgments}
      The research was partly funded by the Icelandic Natural Science Foundation,
      the University of Iceland Research Fund, the National Science 
      Council of Taiwan under Grant No.\ 91-2119-M-007-004, and the 
      National Center for Theoretical Sciences, Tsing Hua University, Hsinchu 
      Taiwan. VG acknowledges instructive discussions with Llorens Serra and 
      a beneficial exchange of programming ideas with Ingibj{\"o}rg 
      Magn{\'u}sd{\'o}ttir and Gabriel Vasile. 
\end{acknowledgments}

%
%
\bibliographystyle{apsrev}

%
\end{document}